# LNOS - Live Network Operating System

Sajjad Haider [1], Dr. M. M. Yasin [2], Naveed Hussain [1], Muhammad Imran [3]
[1] Department of Information Technology, National University of Modern Languages, Islamabad.
[2] Department of Computer Science, Comsats Institute of Information Technolgoy, Islamabad.
[3] Department of Computer Science, Riphah International University, Islamabad.
{sajjad,naveedhussain}@numl.edu.pk, mmyasin@comsats.edu.pk, imran@riphah.edu.pk

*Abstract – Operating Systems exists since existence of computers, and have been evolving continuously from time to time. In this paper we have reviewed a relatively new or unexplored topic of Live OS. From networking perspective, Live OS is used for establishing Clusters, Firewalls and as Network security assessment tool etc. Our proposed concept is that a Live OS can be established or configured for an organizations specific network requirements with respect to their servers. An important server failure due to hardware or software could take time for remedy of the problem, so for that situation a preconfigured server in the form of Live OS on CD/DVD/USB can be used as an immediate solution. In a network of ten nodes, we stopped the server machine and with necessary adjustments, Live OS replaced the server in less than five minutes.*

*Live OS in a network environment is a quick replacement of the services that are failed due to server failure (hardware or software). It is a cost effective solution for low budget networks. The life of Live OS starts when we boot it from CD/DVD/USB and remains in action for that session. As soon as the machine is rebooted, any work done for that session is gone, (in case we do not store any information on permanent storage media). Live CD/DVD/USB is normally used on systems where we do not have Operating Systems installed. A Live OS can also be used on systems where we already have an installed OS. On the basis of functionality a Live OS can be used for many purposes and has some typical advantages that are not available on other operating systems. Vendors are releasing different distributions of Live OS and is becoming their sole identity in a particular domain like Networks, Security, Education or Entertainment etc. There can be many aspects of Live OS, but Linux based Live OS and their use in the field of networks is the main focus of this paper.*

## I. INTRODUCTION

Networks, especially computer networks rely on the services provided by the Operating System. So, in a network environment, OS becomes the most critical software whose failure leads to disaster in network environment. Ultimately, every type of network relies upon the correct functioning of the OS, as the devices used for establishing, handling and managing networks are OS dependent. So failure of an OS can lead to the failure of a network but the reverse may not be true. Failure of an OS which provides services as a server in a network environment is certainly the failure of the network. So, whenever a Network OS responsible for providing services fails, the network service provided by it are affected and becomes unavailable to the user, and eventually clients become the real victims. For more critical Real-Time environments, where delays or stoppage of services can not be tolerated, different types of solutions are possible, such as RAID supportable systems, multiple processors environment, and other types of redundant hardware modules, including machines and NOS.

Hardware redundancy in the form of machines, RAID supportable systems and multiple processor environments are although expensive, but still widely used, and have special significance in providing uninterruptible services in case of hardware failures for design of graceful degraded environment for an OS to perform the fail soft functionality. A software-based issue is of Operating System, being out of service, due to many reasons. A solution is to prepare a machine again, but it could take a lot of time to reinstall the OS and configure the system to run again on the network for provision of services. A proposed and relatively better and effective solution can be adopted in such a situation that is the use of the Live NOS on a CD/DVD. So, the concept is to design a complete Operating System that can be on a CD/DVD and one can use it without installing it on Hard Disk will provide more flexibility at times, and some specific solutions in case of trouble.

## II. LIVE OS ON CD/DVD/USB

Survival without network is merely impossible just as accessing network and its resources without NOS is impractical. So, Operating System, and Network OS is, and will remain the most essential software for computers. From time to time new ideas have been implemented in Operating Systems in order to improve their efficiency and to make them diverse. Operating Systems now days are comparatively intelligent as compared to the Operating Systems that we have experienced in the past. Operating Systems used now days, either by a home user or by an administrator in a complex network environment for delivery of services have at least one thing in common, that is the network. Operating Systems were initially single user systems and have gone from multi user to network environment. Operating Systems are normally installed on hard drives and the time they require for installation and the space they require depends that how many packages or components we have selected as choice. More selection of packages means, more time will be required and more space will be occupied to store them. So, the OS that we normally used like Windows, Macintosh, Unix and Linux etc are installed on the hard drives. After installation, they normally have their booting files available on the hard drives along with the other relevant data. Whenever a user tries to load the OS, the boot files are loaded from hard drive to memory and OS starts providing the services. Other techniques for booting OS are also possible, and include network booting and booting from a bootable Floppy or CD etc.

A Live OS is an Operating System on CD or DVD-ROM, containing Operating System files, and desired or required

softwares, including Network relevant software like different networking servers i.e. Telnet, Proxy and SSH etc, network security tools like Nessus or ACID, Language Compilers, Office applications and other softwares, all stored on a bootable CD or DVD-ROM that can be directly executed, without installation of the OS or other softwares, on hard drive.

Most Live OS available today are based on the GNU operating system and the Linux kernel. Although Live OS based other operating systems exist, such as Mac OS, MSDOS or Microsoft Windows etc, but their legal status is not yet confirmed and predictable.

### III. USAGE SCENARIOS

The list of Live OS available from the Internet shows the primary functions of different distributions [1]. On the basis of usage and functionality they can be used as:

- Desktop OS
  A GUI based desktop environment containing programs like web browsers and other necessary softwares.
- Educational OS
  Contains educational softwares that are used in education fields.
- Rescue OS
  Includes softwares that are used for the recovery of data when conventional OS cannot access files using its own file system.
- Cluster OS
  Used for establishing a Clustered-computing environment [2, 3].
- Security OS
  As an OS that contains network security and assessment tools like Nmap, Nessus and ACID etc.
- Home Entertainment OS
  More focus is towards home entertainment softwares like audio and video applications.
- Diagnostics OS
  Most of the utilities are related to the diagnosing and testing the hardware.
- Firewall OS
  These types of distributions are used in order to achieve the working of a firewall for the creation of Militarized (MZ) and Demilitarized Zones (DMZ) in the LAN/WAN Environment.
- Forensics OS
  Contains softwares that are used for forensic computer analysis [4, 5].
- Server OS
  Can be used to work as different types of servers i.e. SSH, Telnet and FTP etc
- Educationally customized OS
  Can be customized in an educational domain to supply complete working and studying environment for the students who do not have special skills, as they are unfamiliar with the new technology [6].
- Grid OS
  As installation, configuration and maintenance of Grid services is a difficult task, so a natural extension to the traditional Bootable Cluster CD (BCCD) image is one that focuses on aspects of Grid in the form of (BGCD), i.e. Bootable Grid CD [3].

On the basis of usage scenarios, we see that Live OS actually are Special purpose Live OS [7], ranging from large desktop oriented OS that use compressed file systems to add more and more applications to tiny little CDs that are used to set up small routers or firewalls in embedded environments.

### IV. WORKING DETAILS

Live OS comes to life when it is booted from its CD or DVD. The reason of naming Live OS is because it is "brought to life" upon boot without having to be physically installed on a hard drive [8]. During the boot process the Live OS on CD places its files onto a ram disk in comparison to the other Operating Systems that would normally be installed on a hard drive. Although the technique of using ram disk reduces the RAM available to applications and reduces performance to some extent but even though the benefits achieved by Live OS are fairly large [8]. On using the RAM as ram disk, the memory required for application reduces, despite of this drawback, there are much more benefits that Live OS offers.

Based on the concept of Live OS on CD/DVD and the benefits gained, Live OS on USB is available too, and gaining popularity.

In order to boot a Linux based Live OS, a utility with the name of "syslinux" is used mostly and conforms to the "El Torito" specification which treats a special file on a disk as a floppy diskette image [8]. Other technical details of booting, hardware detection and file systems support in some linux based Live OS are available in [9].

### V. FUTURE OF LIVE OS

Due to the features offered, the Live OS has proved its uniqueness and has given new dimensions to the use of Operating Systems in general. Being equally capable of using in many specialized areas like networks, multimedia, security and education etc, still there are unexplored areas that can gain benefit from the features offered by Live OS.

The idea of Live OS was started in early 1990s by Mac OS 7 [8] and a survey revealed that there are more than 300 Live OSs available in the market and the trend is on the rise. Many linux based Live OS are available on the basis of their functionality. Another important point to consider regarding Live OS is their size. Some special Live distributions are as small in size as 5 MB.

Table1 shows some Linux based Live OS distributions on the basis of their primary functions and small size starting from 5 MB [10].

TABLE I
SMALL SIZE LINUX BASED LIVE OS DISTRIBUTIONS

| # | Name of Live OS | ISO Size (MB) | Primary Function |
|---|---|---|---|
| 1 | GeeXboX | 5 | Home Entertainment |
| 2 | CHAOS | 8 | Clustering |
| 3 | Thinstation | 9 | Thin Client |
| 4 | Rxlinux | 10 | Server |
| 5 | Repairlix | 11 | Rescue |
| 6 | Trinux | 19 | Security |

Live OSs that are relatively bigger in size and cannot be stored on smaller USB devices or CDs, are stored on DVD. There are large sized general purpose Live OS available to be used from CD or DVD [10] and some of them are shown in Table II.

TABLE II
LARGE SIZE LINUX BASED LIVE OS DISTRIBUTIONS

| # | Name of Live OS | ISO Size (MB) | Primary Function |
|---|---|---|---|
| 1 | PaiPix | 1720 | Science |
| 2 | Suse Live-Eval | 1446 | Desktop |
| 3 | Knoppix | 700 | OS Replacement |
| 4 | Burnix | 690 | Clustering |
| 5 | Plan-B | 658 | Forensics & Rescue |
| 6 | BackTrack | 625 | Security |

## VI. LIVE NETWORK OS: PROPOSED WORK

Based on the above discussion and the features offered by Live OS, a customized Live Network OS can be established for satisfying the needs of a network environment. A complex network environment is a combination of many types of servers, and different types of servers have different types of configurations. When a network server fails due to any reason, then reconfiguring it, or due to OS failure, reinstalling it, would be a time consuming task that perhaps would not always be affordable. So, our idea is to establish Live Network OS, so that in case of a failure, i.e. Hardware, OS or Server failure, instead of wasting time in installing OS from scratch or start configuring or installing the servers again, we should use Live Network OS.

As many Live OS distributions, specially Linux based, are becoming popular due to their primary function, so the idea of a customized or special server software, designed for a particular organization, can also be configured on a Live OS. The benefit would be that the time for installing the server again, and then configuring it would be saved as it would already be installed and configured on a Live Network OS, and it would yet be another Live OS distribution, used specifically in the scenarios for which it was build.

Proposed work is suitable for establishing servers like anonymous ftp server containing material for download, or secure shell or telnet server for using language compilers available in operating system.

Proposed system will not work for some specialized servers like mail servers. The reason is that mails need to be stored and spooled, while the proposed system can not store information in case of CD or DVD.

In order to check the working of this idea, a network environment of ten machines was created. One of the machines out of them was Anonymous FTP and Telnet server. FTP and Telnet servers were stopped and another machine with Live OS was started and configured as Anonymous FTP and Telnet server. Name server entries that were previously providing old machine name was changed to provide naming information of Live OS. Whole process took less than 5 minutes that could never have been possible otherwise.

## VII. CONCLUSION

Conclusion of this paper is that Live OS can be used in a specific network environment and would always add or would always be seen as an improvement in services provided by that network.

Live OS is found to be a cost effective solution for a network environment that has limited budget allocations and do not have immediate replacement of the failed device.

Customized client/server environment can be established and used in Live OS, and another important benefit for this scheme would be its mobility. As it would be installed and configured on CD/DVD etc, it would be portable in any network environment that uses the same hardware for which Live Network OS was primarily designed.

ACKNOWLEDGMENTS

We are thankful to all of our colleagues including Mr. Abdul Rauf, Mr. Muhammad. Affan, Mr. Fahad Muqaddas, Mr. Raza Pervaiz and Mr. Muhammad Aqeel who have been very helpful and thoughtful while delivering their comments and valuable suggestions.